# High-z gamma-ray bursts for unraveling the dark ages mission HiZ-GUNDAM

Daisuke Yonetoku[a], Tatehiro Mihara[b], Tatsuya Sawano[a], Hirokazu Ikeda[c], Atsushi Harayama[c], Shunsuke Takata[a], Kazuki Yoshida[a], Hiroki Seta[a], Asuka Toyanago[a], Yasuaki Kagawa[a], Kentaro Kawai[a], Nobuyuki Kawai[d], Takanori Sakamoto[e], Motoko Serino[b], Shunsuke Kurosawa[f], Shuichi Gunji[g], Toru Tanimori[h], Toshio Murakami[a], Yoichi Yatsu[d], Kazutaka Yamaoka[i], Atsumasa Yoshida[e], Koji Kawabata[j], Toshio Matsumoto[k], Koji Tsumura[f], Shuji Matsuura[c], Mai Shirahata[l], Hirofumi Okita[l], Kensi Yanagisawa[l], Michitoshi Yoshida[j], Kentaro Motohara[l], and HiZ-GUNDAM working group

[a]Kanazawa University, Kakuma, Kanazawa, Ishikawa 920-1192, Japan
[b]RIKEN, 2-1, Hirosawa, Wako, Saitama 351-0198, Japan
[c]ISAS/JAXA, 3-1-1, Yoshinodai, Chuo-ku, Sagamihara, Kanagawa 252-5210, Japan
[d]Tokyo Institute of Technology, 2-12-1, Ookayama, Meguro-ku, Tokyo 152-8550, Japan
[e]Aoyama Gakuin University, 5-10-1, Fuchinobe, Chuo-ku, Sagamihara, Kanagawa, 252-5258
[f]Tohoku University, 6-6, Aoba, Aramaki-aza, Aoba-ku, Sendai, Miyagi 980-8579, Japan
[g]Yamagata University, 1-4-12 Koshirakawa-machi, Yamagata, Yamagata 990-8506, Japan
[h]Kyoto University, Oiwake-cho, Kitashirakawa, Kyoto 606-8502, Japan
[i]Nagoya University, Furo-cho, Chikusa-ku, Nagoya, 464-8601, Japan
[j]Hiroshima University, 1-3-1, Kagamiyama, Higashi-Hiroshima, 739-8526, Japan
[k]Academia Sinica, No.1, Roosevelt Rd, Sec. 4 Taipei 10617, Taiwan
[l]Institute of Astronomy, University of Tokyo 2-21-1 Osawa, Mitaka, Tokyo 181-0015, Japan

**ABSTRACT**

We are now investigating and studying a small satellite mission HiZ-GUNDAM for future observation of gamma-ray bursts (GRBs). The mission concept is to probe "the end of dark ages and the dawn of formation of astronomical objects", i.e. the physical condition of early universe beyond the redshift $z > 7$. We will consider two kinds of mission payloads, (1) wide field X-ray imaging detectors for GRB discovery, and (2) a near infrared telescope with 30 cm in diameter to select the high-z GRB candidates effectively. In this paper, we explain some requirements to promote the GRB cosmology based on the past observations, and also introduce the mission concept of HiZ-GUNDAM and basic development of X-ray imaging detectors.

## 1. INTRODUCTION

Gamma-Ray Bursts (GRBs) are the most energetic explosions in the early universe. Most of all GRBs occur at the cosmological distance, and their average redshift is $z \sim 2$ in the Swift era. GRBs are thought to be one of the most promising tools to probe the early universe. The remote distance record of GRBs rapidly grows since the discovery of afterglow phenomena in 1997, compared with quasars, normal galaxies and supernova explosions. The GRB distance record is $z = 8.26$ of GRB 090423 ($z = 9.4$ of GRB 090429 by photometric observation), and we expect to make a new record beyond the redshift $z > 10$ in near future. However, for those high redshift (high-z) GRB observations, unfortunately, information about physical condition in the early universe has not been obtained. As a next step, we hope to probe "the end of dark ages and the dawn of formation of astronomical objects" using high-z GRBs, especially for the history of first star formation, cosmic reionization and nucleosynthesis of heavy elements.

In general, the brightness of optical/near infrared afterglow rapidly decreases in time (roughly proportional to $t^{-1}$). An essential key to the observational GRB cosmology is how to realize the follow-up spectroscopic observations with large telescopes during the afterglow is still bright. On the other hand, limited observation times of large telescopes cannot allow us to perform the follow-up observations for every GRB. Therefore we need to select the high-z GRB candidate with near infrared observation as early as possible since the GRB trigger.

We organized a working group to investigate and study a future satellite mission to pioneer the frontier of GRB cosmology, named HiZ-GUNDAM (High-z gamma-ray bursts for unraveling the dark ages mission). Mission payloads of this satellite are wide field X-ray imaging detector to find prompt emission of GRBs, and near infrared telescope with 30 cm in diameter to find the high-z GRB candidate. JANUS[1][2][3] and Lobster[4] are the satellite missions with the same scientific objectives, and they are proposed to NASA's mission. The total weight of HiZ-GUNDAM including both mission payloads and satellite bas is ~300 kg which is rather smaller than JANUS and Lobster, but we carefully designed to satisfy minimum requirements. In this paper, we introduce the HiZ-GUNDAM mission and basic development of mission payload, especially for the X-ray imaging detector.

## 2. MISSION REQUIRMENT

In this section, we summarize past observation results for both prompt emissions of GRBs and following afterglows, and consider possible requirement for future observation in view of GRB cosmology.

### 2.1 Wide Field X-ray Imaging Detector

In figure 1 (left), we show total time duration of prompt emission of GRBs detected by *Swift*/BAT[5][6] as a function of redshift. The time duration of high-z GRBs should be longer because of cosmological time dilation, but the highest 3 events show quite short time duration of $T_{90} < 10$ sec (intrinsic duration is less than 2 seconds in the rest frame of GRBs). At present, there is no events in the red square region while we expect such GRBs should exist. In figure 1 (right), we show a 1 second peak photon flux as a function of redshift. The red solid line is an equivalent line of photon luminosity, and blue dashed line is an effective sensitivity of *Swift*/BAT rate trigger. We can recognize that the 3 highest redshift events are extremely bright events, and their apparent brightness is comparable to the detection threshold of *Swift*/BAT. According to these figures, for high-z GRBs, we can say that we observed only a small fraction (the brightest part) of entire emission, and large amount of emission could not be detected.

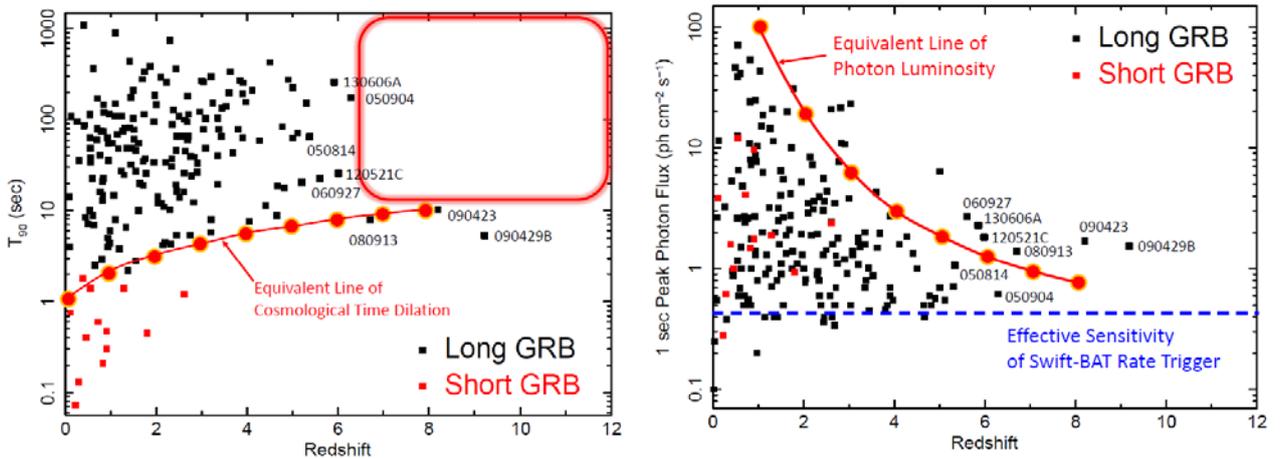

Figure 1. (Left) The time duration ($T_{90}$) of prompt emission of GRBs as a function of redshift. The 3 highest redshift events show shorter time duration as we expect, and there is no events in the region shown as red box at present while this is expected because of the cosmological time dilation. (Right) A 1 second peak photon flux as a function of redshift. The blue dashed line is an effective sensitivity of *Swift*/BAT rate trigger. The 3 highest events are extremely bright events compared with the equivalent line of photon luminosity.

According to the past observations, the peak energy (Ep: typical energy which corresponds to the energy at the maximum flux of $\nu F_\nu$ spectrum) of high-z GRBs are rather small, for example, Ep = 42.1 ± 5.6 keV[7] (GRB 090429B at z = 9.4[8]), Ep = 54 ± 22 keV[9] (GRB 090423 at z = 8.26[10]), Ep = 26.7 ± 4.4 keV[11] (GRB 120521C at z = 6.0[12]), and Ep = 22.6 ± 0.3 keV[13] (GRB 060927 at z = 5.47[14]) while there are exceptions. Therefore, it is important to extend the observation range toward lower energy band, i.e. E < 10 keV, rather than higher energy of E > 150 keV. Additionally, an imaging trigger function of *Swift*/BAT (method to find a transient with subtraction of sky images obtained by longer exposures) effectively improve its detection sensitivity. Therefore, it is important to realize the imaging trigger in soft X-ray band of E < 10 keV.

The event rate of high-z GRBs, which is strongly connected with the star formation rate in the early universe, is not known and one of the top science theme of HiZ-GUNDAM. According to some theoretical works based on the past observations[15], an expected event rate of high-z GRB may ~10 events/yr/str for z > 7 while this value has large uncertainty. At least, *Swift*/BAT finds high-z GRBs with the rate of ~1 event/yr. If we make alerts of high-z GRBs to encourage the follow-up observation by large scale telescopes, we need the wide field of view of ~1 steradian.

We calculated the sensitivity of wide field X-ray imaging detector. In this calculation, we consider the cosmic X-ray background (CXB) and/or the galactic ridge X-ray emission as the diffuse background. We did not include non-X-ray background from surrounding materials and also the contamination of point sources, because they depend on the design of instruments and the operation attitude, and so on. We used the functional form of CXB[16] and one of the galactic ridge X-ray emission[17], and we estimated the net X-ray count rates in 1 – 20 keV as 13.5 photons/cm$^2$/s/str and 55.6 photons/cm$^2$/s, respectively. The gamma-ray spectrum of GRBs are well described by a smoothly broken power-law[18], and the typical photon index of low- and high-energy parts are -1 and -2.5[19], respectively.

In figure 2 (left), we show the trigger sensitivities for the photon flux with 8σ significance as a function of peak energy of GRB's prompt emission. In these calculations, we assume the field of view of 1 steradian, and the energy range of 1 – 10 keV. We show two cases of background conditions, CXB only (high galactic latitude – red solid line) and CXB with Galactic Ridge emission (low galactic latitude – blue solid line). For easy comparison, we also show the trigger sensitivity of Swift-BAT instruments[20] (effective energy range of 15 – 150 keV and detecting area of 5200 cm$^2$. As shown in figure 2 (left), even if the effective area of X-ray imaging detector is smaller than the one of Swift/BAT, it has better sensitivity for X-ray transients with main emission in E < 10 keV, for example high-z GRBs, X-ray flash, supernova shock breakouts, tidal disruption events around super massive black hole, and so on. Moreover, as shown in figure 2 (right), the wide field X-ray imaging detector in the energy range of E < 10 keV with the effective area larger than 100 cm$^2$ is one of the discovery space for X-ray astrophysics. We can consider several configurations of detector setting, so we will investigate how to optimize the sensitivity for the wide field observations in soft X-ray band.

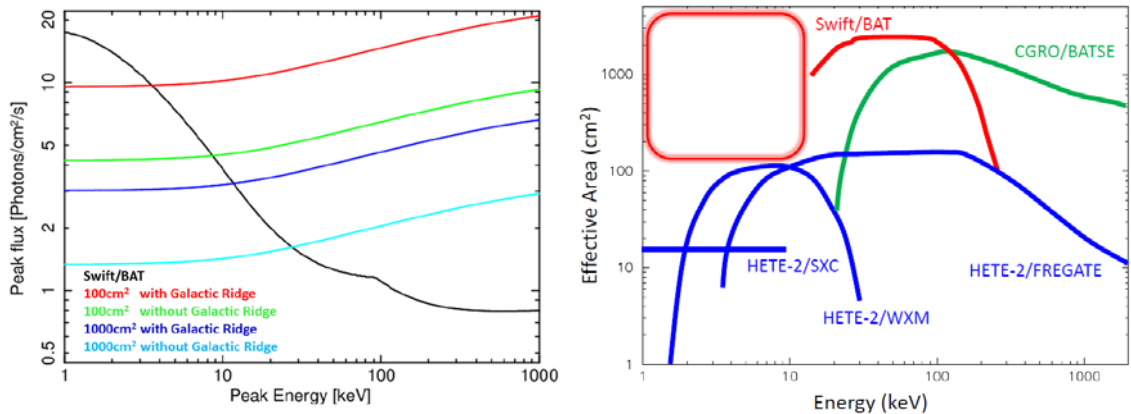

Figure 2. (Left) Sensitivity curves with 8σ confidence level as a function of peak energy. The black solid line is one of *Swift*/BAT[20]. The red and green lines are ones in the case of 100 cm$^2$ detector area with/without galactic ridge X-ray emission, respectively. The blue and cyan lines are the same but for 1000 cm$^2$ cases. (Right) Effective areas of CGRO/BATSE (green), HETE-2 (blue), and Swift/BAT (red). The region shown by red square (energy range of less than 10 keV, and effective area of larger than 100 cm$^2$) is a possible frontier of wide field X-ray observation.

## 2.2 Near Infrared Telescope

Once GRBs are detected, HiZ-GUNDAM automatically changes its attitude and starts follow-up observations like Swift. In the early universe, for example the age of first star generation, most of all baryons were Hydrogen or Helium, and abundances of the other heavy elements were much lower than the present universe. Especially the Lyα emission and/or Lyα break structures are the most important signature. If we focus the redshift of $z > 7$ (and $z > 10$), we have to observe GRBs in near infrared (NIR) band of the wavelength enough longer than $\lambda > 0.97$ μm (and $\lambda > 1.34$ μm).

Kann et al. (2011)[21] summarized a lot of past observations of early-to-late afterglows (see figure 5 of their paper). Referring their results, in figure 3, we converted them into the expected lightcurves if all GRBs locate at $z = 7$. Then we consider following four points. (1) A time axis is stretched by a factor of 4 because of cosmological time dilation. (2) The luminosity distances are 6.51 Gpc and 68.0 Gpc for $z = 1$ and 7, respectively. (3) A flux density $F_\lambda$ is proportional to $\lambda^{-1.4}$ as an average property of afterglow spectrum. (4) The difference of center wavelength between Rc (0.6588 μm) and J (1.215 μm) makes a conversion coefficients of $(1.215/0.6588)^{-1.4} = 0.424 \sim +0.9$ mag(AB). Finally we estimated that the flux density of GRBs at $z = 7$ observed in J band is averagely +3.9 mag(AB) dimmer than ones at $z = 1$ observed in Rc band (compared with figure 5 of Kann et al. 2011[21]).

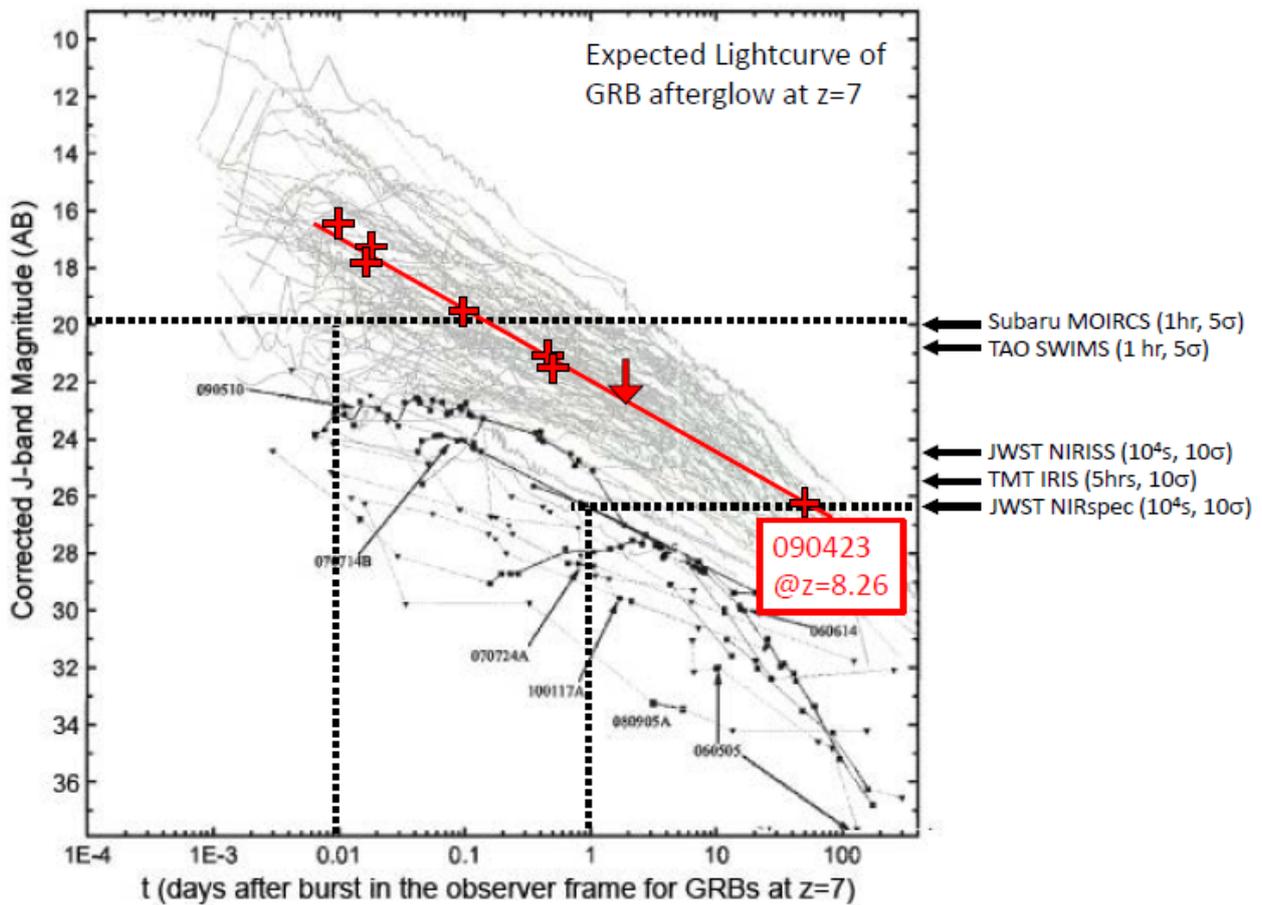

Figure 3. An expected lightcurve of GRB afterglows at $z = 7$. This figure is converted from the one of Kann et al. (2011)[21]. The red crosses are observations of GRB 090423 at $z = 8.26$. The spectroscopic sensitivity of Subaru, TAO, JWST, TMT are also shown at the outside of the panel. If we send an alert message for the discovery of high-z GRB candidates until 0.01 days since GRB trigger, large observatories can perform the spectroscopic observations.

On the other hand, we hope the large telescopes to perform the spectroscopic observation with their full capability. Then the sensitivity of spectroscopic observation is 20.1 mag(AB) for Subaru/MOIRCS (R = 700, J band, 1 hour exposure, 5σ)[22], ~21.9 mag(AB) for TAO SWIMS (R = 1,000, J band, 1 hour exposure, 5σ)[23][24][25], ~24 mag(AB) for JWST NIRISS (slitless R = 150, at ~1 μm, $10^4$ sec exposure, 10σ)[26], ~25.5 mag(AB) for TMT IRIS (R = 4,000, J band, 5 hours exposure, 10σ)[27]. When we assume the Subaru's and TAO's response time for ToO (Time of Opportunity) observations as 1~24 hours, we need measure rough redshifts, for example 5 < z < 7 and z > 7, with photometric observation, and also to send alert message with the redshift information until 0.01 days since GRB trigger to cover almost all samples. Then we need the sensitivity of ~20 mag(AB) for a few minutes exposure in both optical and near infrared bands to measure Lyα break of Hydrogen. These are the requirements for our NIR telescope. JWST and TMT may enable to perform the ToO spectroscopic observation after 1~10 days since GRB trigger. Then, the requirement for time limitation of GRB alert will be loose, and we may accept it after ~1day since GRB trigger. In this case, we need much more sensitivity for NIR telescope, for example ~24 mag(AB). On the other hand, the requirement of 20 mag(AB) at 0.01 days is acceptable for JWST and TMT.

## 3. WIDE FIELD X-RAY IMAGING DETECTOR

As summarized in the previous section, we need wide field X-ray imaging detector to cover the energy range of E < 10 keV with 1 steradian field of view. A localization accuracy for prompt emission of GRBs is needed to perform the follow-up observation with the NIR telescope and ground based/space observatories, so its capability of 10 arcmin are required. There are several wide field X-ray imaging instruments, for example coded aperture mask system, Lobster-eye optics, slit camera and so on. Based on the requirement shown in the previous section and also considering limited resources of small satellite, we selected 1-dimensional X-ray imaging detectors with coded aperture mask system at present. In this section, we introduce basic developments for 1-dimensional silicon strip detectors and readout electronics.

### 3.1 Silicon Strip Detectors (SSDs)

To understand the difference of performance caused by the several designs of electrode, we developed several types of SSDs listed in table 1. In figure 4 (left), we show 6 kinds of 1-dimensional silicon strip detectors (SSDs) processed by Hamamatsu Photonics. All designs have 500 μm thickness (the ratio of photo electric absorption is 98 % for 10 keV X-ray photons) and 64 electrode channels with 300 μm pitch length. The top and bottom 3 pieces in figure 4 (left) have 32 mm (SSD-L) and 16 mm (SSD-S) strip length, respectively. The left, middle and right pieces have different electrode width of 100 μm, 200 μm and 280 μm, respectively. All designs have fan-out structures with the pad pitch of 91.2 μm for a purpose of direct connection between the SSD and readout ASICs. The figure 4 (right) is an enlarged photograph of the fan-out structure. The backside of SSD is the cathode of PIN diode (pulse high voltage side), and this +HV appears surrounding outside-edge of front surface. To reduce the leakage current between this HV area and the most outside strip, we prepared guard-ring structure which surrounds all 64 strips. We investigated some fundamental characteristics for each SSDs, and the test results are listed in table 1.

We measured the leakage current as a function of reverse bias voltage. At the temperature of 20 degrees, the average leakage current of each strip of SSD-L-280 μm model is 710 pA, and the standard deviation of 37 pA. Therefore the characteristics of leakage current is well ordered in 5 % level. In table 1, we show typical leakage current for each design at the reverse bias of +200 V and the temperature of 0 degrees. In this experiment, we measured the leakage current until the reverse bias of 300 V, and we found the both designs of 100 μm electrode tend to show a breakdown compared with 280 μm and 200 μm design. This can be recognized that the electric fields inside the SSD are strongly bended around the anode strips because of their wider separation, and the lines of electric field concentrate at the edges of each strip.

These SSDs have two kinds of capacitances, i.e. a body capacitance between the cathode and anode of each strip, and an inter-strip capacitance between neighboring anode strips. We separately measured these characteristics. For the body capacitance, we connected all 64 channels anodes into one, and measured the capacitance between the cathode and the combined anode as a function of reverse voltage. These capacitances can be basically described as a power-law of $V^{-0.5}$, which is typical characteristics of semiconductor devices, before it achieves a full depletion. In table 1, we show the body capacitance per strip and the reverse voltage at the full depletion.

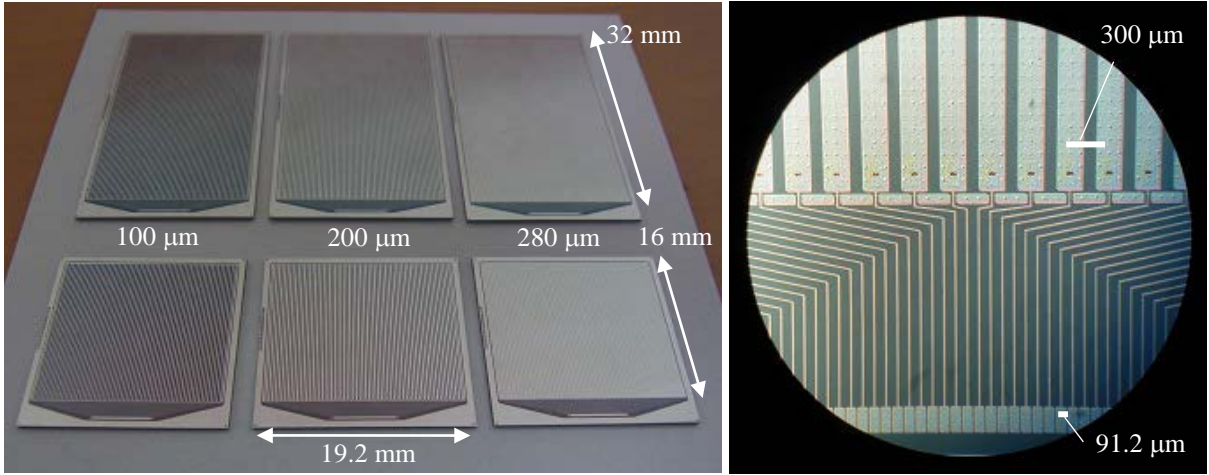

Figure 4. The left panel shows silicon strip detectors (SSDs) for different designs. Each device contains 64 strips with 300 μm pitch width, and the width of electrode is 100 μm (left), 200 μm (center) and 280 μm (right), respectively. The fan-out structures of electrode are prepared for the purpose of direct connection between SSD and readout ASIC. The right panel is a close-up photo of the fan-out structure.

Table 1. Configurations of SSDs

|  | SSD-S (16 mm) | | | SSD-L (32 mm) | | |
|---|---|---|---|---|---|---|
| Detector size | 21.1 mm x 19.7 mm | | | 21.1 mm x 35.7 mm | | |
| Effective area | 19.2 mm x 16.0 mm | | | 19.2 mm x 32.0 mm | | |
| Thickness | 500 μm | | | 500 μm | | |
| Number of strip | 64 | | | 64 | | |
| Strip pitch | 300 μm | | | 300 μm | | |
| Readout pad pitch (fan-out structure) | 91.2 μm | | | 91.2 μm | | |
| Electrode width | 100 μm | 200 μm | 280 μm | 100 μm | 200 μm | 280 μm |
| Leakage current (pA) @ 0 deg, 200 V | – [*1] | 35.5 | 35.3 | – [*1] | 70.9 | 86.1 |
| Breakdown voltage (V) | 60 | > 300 | > 300 | 60 | > 300 | > 300 |
| Reverse bias at full depletion (V) | 94.0 | 76.2 | 73.2 | 103.4 | 83.9 | 83.3 |
| Body capacitance (pF/strip) | 1.10 | 1.16 | 1.15 | 1.96 | 2.20 | 2.27 |
| Inter-strip capacitance (pF/strip) | 2.5 | 3.7 | 5.6 | 4.4 | 6.8 | 10.0 |

*1. Already achieved the breakdown voltage.

Next, we also measured the inter-strip capacitance for all models. Then we notice one central strip and the other 63 channels are combined, and we measured the capacitance between them applying the reverse bias voltages. In this measurement, the inter-strip capacitance includes not only the neighboring strips but also the other remote strips. These inter-strip capacitances rapidly decrease as a function of reverse biases, and achieve a steady state around 70 V for all designs. This can be interpreted as the depletion layer close to cathode electrodes grows enough until this reverse voltages, and after that it grows toward the depth direction until the full depletion.

Combined with 1-dimensional coded aperture mask, we can perform the X-ray imaging observation with wide field of view. When we set the coded aperture mask with $d_1$ = 300 μm pitches (same of SSD strip $d_2$ = 300 μm) at D = 20 cm distance from the detector plane, we can realize the localization accuracy of $\theta \sim \tan^{-1}(\sqrt{d_1^2+d_2^2}/D) \sim 7.3$ arcmin by the geometrical condition. According to the Swift observations, for bright GRBs and X-ray transients, this accuracy can be improved to ~ 4 arcmin because of a weighting factor of photon statistics.

## 3.2 Readout ASIC (ALEX-01)

One of the key technologies for X-ray imaging detector is multi-channel readout methods. We are developing a readout analog ASIC (application specific integrated circuit) for the SSDs supported by "Open IP project" promoted by ISAS/JAXA. This ASIC, named ALEX-01 (ASICS for low energy X-ray ver. 01), is fabricated by X-FAB XH-035 (0.35 μm CMOS) processes, and is specifically designed to readout small charge signals from X-ray with 1 – 20 keV. Therefore ALEX-01 has a quite larger electric gain compared with our ASIC series for example (KW series by ISAS/JAXA)[28][29][30][31][32][33]. We show the ALEX-01 mounted on an IC package of QC-160360-WZ (Kyocera) and enlarged photograph in figure 5 (left) and (right), respectively. This ALEX-01 is just a test piece, so only 40 channels of analog inputs are connected to the package, and the other 24 channels are not used.

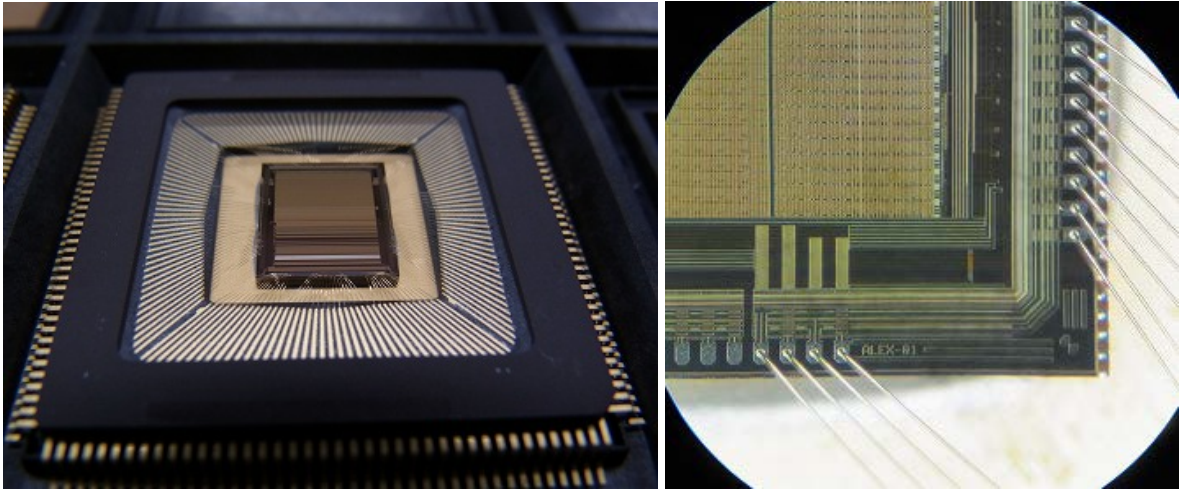

Figure 5. (Left) A photo of readout ASIC named ALEX-01 (ASICS for Low-Energy X-ray ver. 01) mounted on a general IC package for test usage. (Right) A close-up photo of ALEX-01. The pad pitch of analog input is 91.2 μm as the same pitch of fan-out structure of SSDs.

ALEX-01 has general readout scheme for X-ray and gamma-ray instruments, i.e. preamplifier, shaping (filtering) amplifier, sample hold, and Wilkinson-type analog to digital converter (A/D converter), as shown in figure 6. The preamplifier has large gain with the feedback capacitance of 0.016 pF (0.032 pF) and the time constant of ~700 μsec, whose gain can be changed by MOS switches keeping the same time constant by a register setting inside the ASIC. We prepared two kinds of shaping amplifiers. One has a fast peaking time of ~0.8 μsec to create trigger signal, and the other has a slow peaking time of ~4 μsec to be readout with high signal to noise ratio. Once the trigger signal is thrown to the following FPGA with a clock of 40 MHz, it sends the return signal after a delay of 3.75 μsec to switch off a part of circuit line in the ASIC, and hold the charge signal (pulse height) of the slow-shaping amplifier output. These configurations are summarized in table 2.

Simultaneously, the FPGA sends another request-signal to the ASIC to make a ramp wave for A/D conversion. A comparator in the ASIC, which has two voltage inputs from the sample-held slow-shaping amplifier and the ramp wave, is connected to a 10-bit counter. We get the digitized pulse height by measuring the time duration how long the comparator's output keeps high level. For adjustment of the input range, the counter refers an external clock and the ramp wave has a 4-bit variable slope. The sample-hold data are encoded and sent to the FPGA with serial data transmission. We can selectively change data transfer formats, i.e. we can send all data of 64 channels once any channels create trigger signals, or only the data with a pulse height above either analog or digital threshold we set.

The ASIC has one common control register to set the input capacitance of the preamplifier, the ramp speed, data transfer criteria, and so on, and 64 local control registers for the individual channels to set the digital threshold, offset level of the shaping amplifier, veto and so on. Each register has a copy register to monitor if there is inconsistency between the values of the two registers, which enable us to detect single-event upsets of these registers.

Table 2. Specifications of ALEX-01

| Chip name | ALEX-01 (ASICS for low energy X-ray ver. 01) |
|---|---|
| Chip size | 8.800 mm x 7.460 mm |
| No. of channels | 64 |
| Pad pitch of analog input | 91.2 μm |
| Fabrication and process | X-FAB XH-035 (0.35 μm CMOS) |
| Gain of pre-amplifier | 55 mV/fC (@ $C_f$ = 0.016 pF) |
| Time constant of pre-amplifier | 700 μsec |
| Total gain | 750 mV/fC |
| Peaking time of fast/slow amplifier | ~0.8 μsec (fast) & ~4 μsec (slow) |
| Power rail | ±1.65 V (analog and digital), +3.3 V (digital) |
| Power consumption | ~120 mW |
| Dynamic range | 6300 $e^-$ (23 keV for Si) |
| Typical noise level ( no-load) | 88 $e^-$ |

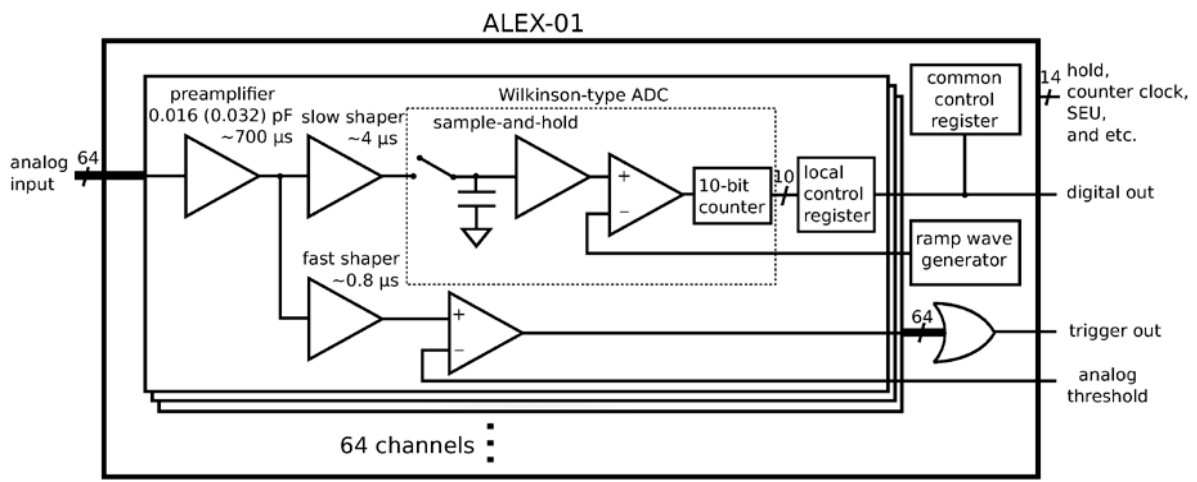

Figure 6. Schematic view of signal processing by ASIC ALEX-01.

### 3.3 Characteristics of ALEX-01

We investigated characteristics of ALEX-01. Here we introduce a linearity for input signal, a capacitance characteristic for the analog input, and a noise characteristics. In figure 7 (left), we show the linearity of output pulse height (ADC) as a function of input signal from function generator. Here we set a voltage output from the function generator as constant, and change the attenuation level to create various input signals. As shown the figure, there is a discontinuous step at E = 10 keV, but this is caused by imperfect calibrations of two kinds of attenuator. We can say that the linearity of ALEX-01 is ordered within 2 % deviations between the energy ranges of 1 – 23 keV (Probably less than 0.5 % without the discontinuities).

In figure 7 (right), we show a noise characteristics as a function of input capacitance. We connected some ceramic capacitance on the analog input line of ALEX-01. Then we measured a Gaussian distribution of pulse height of input test pulse, and estimate the equivalent number of electrons as a function of value of capacitance. The best fit line shows that the gradient of the slope is ~3 $e^-$/pF. In this experiment, there are some additional capacitance from a bonding wire and an electrode of IC packages. Therefore an intercept of the line (at 0 pF) is somewhat larger value rather than we expected. When we measure the noise level for non-connected channels of ALEX-01, we obtained 88 $e^-$ for no-load condition, and this is almost consistent with a result by numerical simulations with T-SPICE simulator. Therefore we conclude the performance of ALEX-01 itself almost achieves the design value.

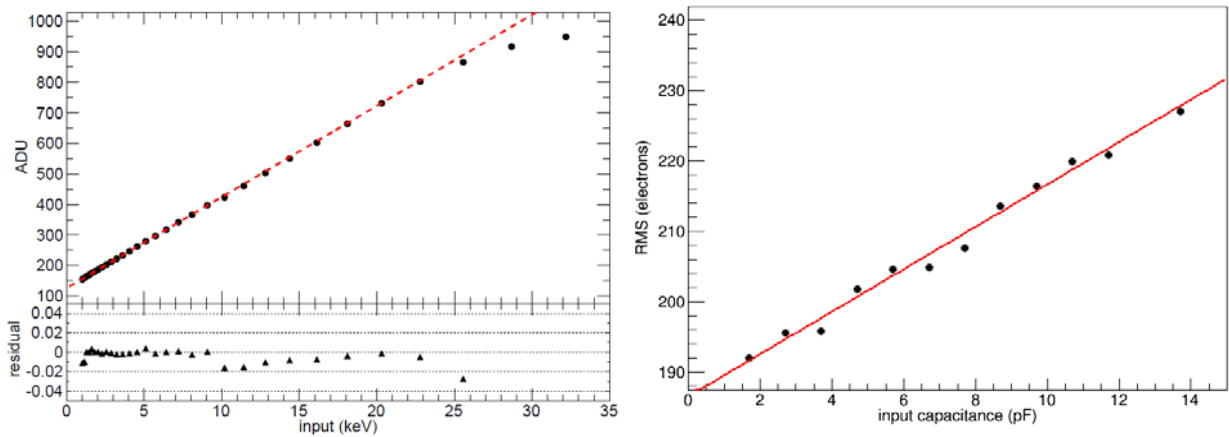

Figure 7. (Left) A linearity of output pulse height (ADU) as a function of input charges (converted to equivalent energy). The bottom panel is a residual between the data and the best fit linear function for E < 10 keV. Although we can see a discontinuous step at E = 10 keV, this is caused by switching of two kinds of attenuator. (Right) A noise characteristics as a function of input capacitance. The slope can be described as ~3 e-/pF.

## 4. NEAR INFRARED (NIR) TELESCOPE

Responding to new GRB discoveries with the wide field X-ray instruments, HiZ-GUNDAM automatically start to the follow-up observation with the NIR telescope. In this section, we introduce a specification of IR telescope. This design of telescope is very similar to the one of CIBER-2[34] sounding rocket experiment for the cosmic infrared background observations (see also CIBER project[35]).

According to the requirement discussed in section 2, we consider the NIR telescope with an aperture size of 30 cm in diameter to realize the limiting magnitude of ~20 mag(AB). The NIR telescope should perform pointing observations for GRBs occurring in the large field (~1 steradian) of view of X-ray imaging detector. Therefore we add a large aperture shade (reflection baffle) in front of telescope to avoid any inflow of optical/NIR radiation from the sun and the earth. The size of aperture shade is 68 cm in diameter with the opening half angle of ~30 degrees (see figure 8. left). To do so, HiZ-GUNDAM can realize extremely large sky coverage, for example 30 degree for the solar side, 56 degree for the earth side (depends on the orbital altitude, e.g. 600 km). This aperture shade is made of two materials, i.e. a PTFE (Teflon) vapor deposition with silver for a reflection side of solar radiation, and an aluminum mirror surface for the other reflection side of radiation from the earth. The PTFE side and an edge of telescope tube work as a radiator of aperture shade and the telescope itself, respectively.

We summarized the configurations of NIR telescope in table 3. At present design, considering the limited size of satellite platform, we will use an offset Gregorian optics as shown in figure 8 (right). This optics has an advantage to reduce some influences of stray light because the ray of light focuses before a secondary mirror. The focal length and F-number are 183.5 cm and F6.1, respectively. We will divide the observation wavelength into 4 band, (1) 0.5 – 0.73 μm, (2) 0.73 – 1.00 μm, (3) 1.00 – 1.30 μm, and (4) 1.30 – 1.70 μm to measure the Lya break structure and determine the rough redshift of z > 7 and z > 10, respectively. If we use HgCdTe (HAWAII array) for NIR band and HiViSI for optical band for the focal plane detector[36][37], the plate size is 2 arcsec/pixel and the field of view is 34 x 34 arcmin$^2$. This design can cover the localization accuracy of X-ray imaging detector (~10 arcmin), and effectively perform the follow-up observations for almost all GRBs. Based on the configuration listed in table 3 and using the zodiacal light spectrum[38], we estimated the sensitivity for each band. Then we assume that the image size of 3x3 pixels, a readout noise of 20 e-, leakage current of detectors of 1 e-/sec, and the throughput efficiency of 50 % including the telescope optics, dichroic mirrors, and detectors. We can recognize that the limiting magnitudes of all bands are satisfied with the requirement shown in section 2.

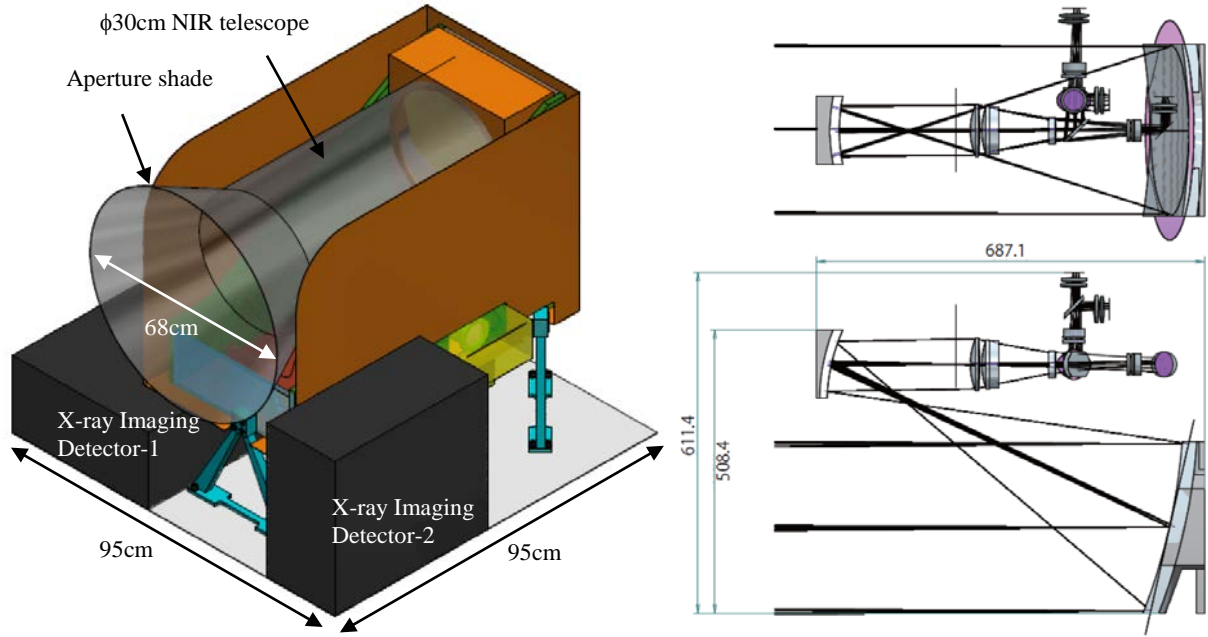

Figure 8. (Left) A simplified structure model of scientific payloads. Two black boxes are the wide field X-ray imaging detector for X- and Y-axis, respectively. (Right) A ray trace simulation with an offset Gregorian optics with 4 kinds of detector modules (2 optical and 2 NIR) on their focal plane.

Table 3. Configurations of NIR telescope

| Aperture size | 30 cm in diameter | | | |
|---|---|---|---|---|
| Optics | Offset Gregorian | | | |
| Focal length (F-number) | 183.5 cm (F6.1) | | | |
| Size of aperture shade | 68 cm in diameter, 30 cm in length | | | |
| Field of view | 34 x 34 arcmin$^2$ | | | |
| Sky coverage | 30 degree for the solar side, 56 degree for the earth side (depends on the orbital altitude) | | | |
| Plate scale | 2.0 arcsec/pixel (1 pixel = 18μm) | | | |
| Operating temperature | < 230 K (for telescope tube), < 100 K (for detectors) | | | |
| Band | 0.5 – 0.73 μm | 0.73 – 1.00 μm | 1.00 – 1.30 μm | 1.30 – 1.70 μm |
| Sensitivity (10σ, 3 min) | 20.5 mag(AB) | 20.3 mag(AB) | 20.1 mag(AB) | 20.1 mag(AB) |
| Detector | HiViSI-Blue | HiViSI-NIR | HgCdTe | HgCdTe |

## 5. ORBIT

The selection of satellite orbit is important for the NIR telescope to keep its operating temperature of T < 230 K. The twilight line of sun synchronous orbit is the best in view of the thermal design for NIR telescope like the one of AKARI satellite[39]. However, in this case, it is difficult to observe the anti-direction of the sun (midnight) which is the best condition for ground based telescope. Therefore we consider the sun synchronous orbit with a local time of 9 hour (21 hour). Then, the thermal condition is worse than the AKARI's orbit[39], but our observation is only in NIR band so this can be acceptable for our requirement of T < 230 K.

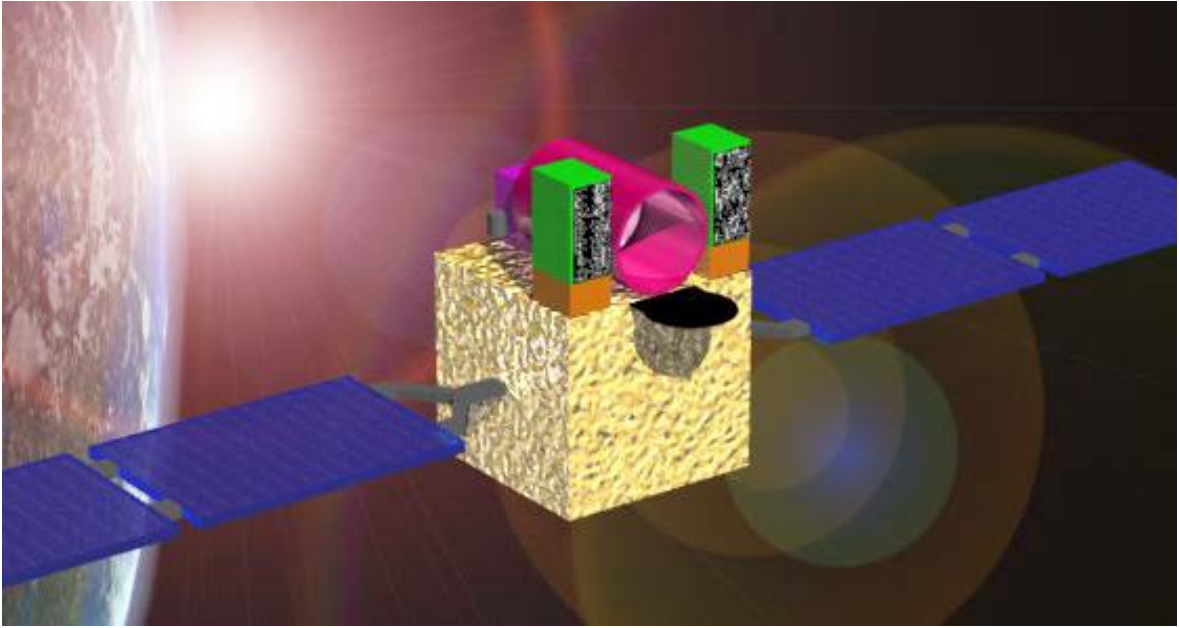

Figure 9. A schematic image of HiZ-GUNDAM satellite.

## 6. SUMMARY

We organized a working group to investigate and study a future satellite mission HiZ-GUNDAM to pioneer the frontier of GRB cosmology (see a schematic view in figure 9). We will install the wide field X-ray imaging detector to observe the energy range of 1 – 10 keV, and the near infrared telescope with 30 cm in diameter. The main aim of HiZ-GUNDAM is to determine the high-z GRB candidate with the photometric observation with NIR telescope. After that, collaborating with the large telescopes, such as Subaru, TAO, JWST and TMT, we hope to reveal the physical condition of the early universe. We hope this mission will strongly promote the high-z observations by GRBs. Moreover, this mission contribute to understand the nature of soft X-ray transients, for example, X-ray Flashes, extended shoft X-ray emissions of short GRBs, shock breakouts from supernova explosions, tidal disruption events of stars falling into black hole, and so on.

## ACKNOWLEDGE

This work is supported in part by the Grant-in-Aid from the Ministry of Educatiion, Culture, Sports, Science and Technology (MEXT) of Japan, No. 25103757, No. 25247038, and Mitsubishi Foundation (DY).